\documentclass[a4paper,11pt]{article}
\pdfoutput=1
\usepackage{jcappub}
\usepackage{inputenc}
\usepackage{mathtools}
\usepackage{xcolor}
\usepackage{float}
\usepackage{orcidlink}
\graphicspath{{plots/}}
\usepackage[makeroom]{cancel}
\usepackage[normalem]{ulem}
\usepackage{slashed,soul,multirow, makecell}

\title{Constraints on the inflationary vacuum and reheating era from NANOGrav}

\author{Debtosh Chowdhury\orcidlink{0000-0002-4302-7356},}
\author{Rounak Nath\orcidlink{0009-0004-0702-4483},}
\author{Sudipta Show\orcidlink{0000-0003-0436-6483}}

\affiliation{Department of Physics, Indian Institute of Technology Kanpur, Kanpur 208016, India}

\emailAdd{debtoshc@iitk.ac.in}
\emailAdd{rounakn24@iitk.ac.in}
\emailAdd{sudiptas@iitk.ac.in}

\abstract{NANOGrav and various pulsar timing array experiments recently reported compelling evidence for a stochastic gravitational wave background (SGWB). Such a background may originate from several astrophysical or cosmological sources. Assuming an inflationary origin, we use the latest NANOGrav 15-year dataset to constrain inflationary parameters, including the tensor spectral index ($n_t$), tensor-to-scalar ratio ($r$), and explore the implications for reheating through constraints on the reheating equation of state ($\omega_{\text{re}}$) and reheating temperature ($T_{\text{re}}$). We find a preference for an extremely blue-tilted tensor spectrum and a non-instantaneous reheating epoch. Despite no concrete evidence on primordial vacua, inflationary vacuum is commonly assumed to be the Bunch-Davies vacuum. In this work, we study modifications to the GW spectrum arising from a two-parameter Bogoliubov family of non-Bunch-Davies vacua. Within this framework, we find that NANOGrav observations favour a subclass of non-Bunch-Davies vacuum, known as the alpha-vacuum. In addition, our analysis demonstrates that the observations strikingly narrow the range of the parameter $\alpha$ that characterizes the vacua. Our analysis indicates that the NANOGrav data can accommodate both matter- and radiation-like reheating scenarios for the standard Bunch-Davies vacuum case. However, within the non-Bunch-Davies framework considered here, a non-matter-like reheating is preferred because matter-like reheating requires relatively large $\alpha$, violating both the NANOGrav upper bound and the backreaction constraint. We further show that a frequency-dependent parametrization of $\alpha$ beyond a threshold frequency can yield a minimal solution that alleviates the blue-tilted issue. Finally, we highlight the possibility of testing such frequency dependence of $\alpha$ through future GW experiments.}
\begin{document} 
\maketitle
\flushbottom
\section{Introduction}

Decades of precise pulsar timing observations from millisecond pulsar arrays provide a platform for detecting the stochastic gravitational wave background (SGWB) in the nHz frequency range. Over the past couple of years, several pulsar timing array (PTA) collaborations, such as NANOGrav~\cite{NANOGrav:2023gor}, EPTA~\cite{EPTA:2023fyk}, IPTA~\cite{Antoniadis:2022pcn}, PPTA~\cite{Zic:2023gta}, InPTA~\cite{Tarafdar:2022toa}, CPTA~\cite{Xu:2023wog} have observed a common red noise spectrum with distinct angular correlations among the pulsars. This correlation is referred to as the Hellings-Downs curve~\cite{Hellings:1983fr}, a hallmark signature of the isotropic stochastic gravitational wave (GW) background. Recently, NANOGrav has reported evidence for the stochastic GW signal at 3.5$\sigma$-4$\sigma$ statistical significance after analysing the 15-year data set comprising observations of 67 pulsars along with their correlations. However, the exact origin of such SGWB remains unsettled as a plethora of cosmological sources such as inflation~\cite{Vagnozzi:2020gtf, Benetti:2021uea, Vagnozzi:2023lwo, Ben-Dayan:2023lwd}, first-order phase transition~\cite{Fujikura:2023lkn, Addazi:2023jvg, Bai:2023cqj, Megias:2023kiy, Han:2023olf, Zu:2023olm, Ghosh:2023aum, Xiao:2023dbb, Li:2023bxy, Gouttenoire:2023bqy, Ahmadvand:2023lpp, Wang:2023bbc, Ellis:2023oxs}, cosmic defects~\cite{Ellis:2023oxs, Ellis:2023tsl, Kitajima:2023vre, Wang:2023len, Lazarides:2023ksx, Eichhorn:2023gat, Kitajima:2023cek, Blasi:2023sej, Gouttenoire:2023ftk, Chowdhury:2023opo, Antusch:2023zjk, Yamada:2023thl, Ge:2023rce, Basilakos:2023xof, Barman:2023fad, Li:2023tdx, Du:2023qvj, Babichev:2023pbf, Gelmini:2023kvo, Zhang:2023nrs}, supermassive black hole binaries~\cite{Ellis:2023oxs, NANOGrav:2023hfp, EPTA:2023xxk, Ellis:2023dgf,Calza:2024qxn} etc. can mimic such SGWB. Later on, in a companion work~\cite{NANOGrav:2023hvm}, the NANOGrav collaboration provided a comprehensive list of cosmological sources that can address the observed data. Among them, one of the intriguing origins of SGWB is the inflationary GW sourced by the tensor perturbations originated during inflation~\cite{PhysRevD.22.1882, Kodama:1984ziu, Malik:2008im, Caprini:2018mtu, Christensen:2018iqi}. 

Inflationary interpretation of SGWB demands a blue-tilted gravitational wave spectrum, i.e., the tensor spectral index, $n_t>0$~\cite{Vagnozzi:2020gtf}. Typically, the GW spectral density depends on two quantities: tensor spectral index ($n_t$) and tensor-to-scalar ratio ($r$). The observed common red noise spectrum is parametrized by the amplitude ($A$) and spectral index ($\gamma$) of the PTA signal, which are related to the tensor spectral index and tensor-to-scalar ratio. As a result, the allowed range of $A$ and $\gamma$ from the observed SGWB can constrain the inflationary parameters as shown in Refs.~\cite{Vagnozzi:2020gtf, Vagnozzi:2023lwo, Ben-Dayan:2023lwd}. It is noteworthy to mention that, although the blue-tilted gravitational wave sourced during inflation explains the NANOGrav observation, it cannot be accommodated within the conventional slow-roll inflationary paradigm, which prefers $n_t<0$ (i.e., red tilted) since $n_t=-r/8$~\cite{Liddle:1993fq}  and $r>0$ from the data~\cite{Planck:2018vyg}.

Besides constraining the inflationary parameters, the NANOGrav observation can also constrain the reheating phenomenon. Usually, in this standard cosmological picture, one assumes that the universe enters the radiation-dominated era right after the end of inflation. In this standard picture, the tensor perturbations at nHz frequencies, a range that is the primary focus of the PTA experiments, re-enter the horizon during the radiation-dominated era, where the equation of state is $1/3$. As a result, the GW relic density depends on $r$ and $n_t$. In case of non-instantaneous reheating~\cite{Kolb:2003ke, Bassett:2005xm, Garcia:2017tuj,  Garcia:2020wiy, Mambrini:2021zpp, Banerjee:2022fiw, Bernal:2022wck, Bernal:2023ura, Chowdhuryand:2024uvi}, the GW energy density depends on the reheating equation-of-state parameter ($\omega_{\text{re}}$) and the reheating temperature ($T_{\text{re}}$). The reheating temperature $T_\text{re}\le$ 50 MeV such that the tensor modes that re-enter the horizon during the reheating phase correspond to nHz frequencies. In addition, the universe had to be radiation-dominated before big bang nucleosynthesis (BBN), which disfavours $T_\text{re}$ below 4 MeV~\cite{Kawasaki:2000en, Hannestad:2004px, deSalas:2015glj, Gerbino:2016sgw}. Thus, the permissible window of reheating temperature becomes quite narrow; in particular, the viable range is approximately 4-50 MeV.

Alongside constraining the reheating era, NANOGrav 15-year data can provide insights into the inflationary vacuum. In standard treatments, the calculation of primordial GWs assumes the inflationary vacuum to be of Bunch-Davies type~\cite{Bunch:1978yq}. However, the choice of the inflationary vacuum is not unique due to the underlying de-Sitter symmetry. So, it is important to investigate how the usual gravitational wave spectrum is modified when the inflationary vacuum is non-Bunch-Davies. In literature, several studies have investigated the effects of non-Bunch-Davies vacuum on the primordial power spectra~\cite{Sriramkumar:2004pj,Ashoorioon:2013eia, Cielo:2024poz, Maity:2025czp,Wood-Saanaoui:2026bma}, non-Gaussianity~\cite{Xue:2008mk,Gong:2013yvl,Aravind:2013lra,Bahrami:2013isa,Kundu:2013gha,Meerburg:2015yka,Ashoorioon:2016lrg,Shukla:2016bnu,Akama:2020jko,Naskar:2020vkd,Kanno:2022mkx,Gong:2023kpe,Ansari:2024pgq,Cielo:2023enz}, and GW spectrum~\cite{Akama:2024vgu,Cielo:2024poz,Fumagalli:2021mpc,Choudhury:2023kam}. Non-Bunch–Davies states have also been explored as a possible way to reconcile inflationary cosmology with the Swampland criteria and the Trans-Planckian
Censorship Conjecture ~\cite{Brahma:2018hrd,Ashoorioon:2018sqb,Brahma:2019unn}. In this work, we compute the GW spectrum for a particular choice of non-Bunch-Davies vacuum, parametrized by two parameters, explore its phenomenological implications, and, for the first time, place observational constraints on the parameters that characterize this vacuum state.

GWs redshift as radiation after production in the early universe and contribute to the radiation energy density as an additional component. Such an extra contribution is parametrized as an excess in the effective number of relativistic species, $\Delta N_{\text{eff}}$. Cosmic microwave background (CMB) experiments measure the quantity $N_{\text{eff}}$ with the following precision $N_{\text{eff}}=2.99\pm0.17$~\cite{Planck:2018vyg}. Such a measured bound on $\Delta N_{\text{eff}}$ puts an upper bound on the energy density stored in the GW parametrized by $\Omega_{\text{GW}}h^2\le 2.4\times 10^{-6}$, irrespective of its origin. The SGWB, having an inflationary origin, has to be consistent with the CMB observations. However, the blue-tilted gravitational wave violates such a constraint, giving rise to the so-called blue-tilted issue of the GW spectrum. However, several solutions have been proposed to address the blue-tilted issue, such as early matter or primordial black hole domination~\cite{Datta:2023xpr, Athron:2024fcj}, running of the tensor spectral index~\cite{Ben-Dayan:2023lwd}, inclusion of the contribution from astrophysical sources of GWs~\cite{Ben-Dayan:2023lwd}, etc.

In this work, we use the NANOGrav 15-year data to probe the inflationary reheating phase, considering that the observed SGWB has an inflationary origin. In this case, one can express amplitude $A$ and spectral index $\gamma$ of the GW spectrum as functions of the inflationary parameters $r$, $n_t$, and the reheating parameters $\omega_{\text{re}}$,~$T_{\text{re}}$. We perform a Markov chain Monte Carlo (MCMC) analysis to determine the range of these parameters consistent with the NANOGrav data. Next, we calculate the gravitational wave spectrum considering a two-parameter Bogoliubov family of non-Bunch-Davies inflationary vacuum parametrized by $\alpha_t$ and $\beta_t$. In this scenario, the tensor power spectrum depends explicitly on the vacuum parameters $\alpha_t$ and $\beta_t$, and consequently, so does the GW spectrum. The smallness of the amplitude of the power spectrum constrains the parameter $\beta_t$ to be 0. We perform an additional MCMC analysis to constrain the vacuum parameter $\alpha_t$, thereby restricting its allowed range. Finally, we show that introducing a scale-dependency on the vacuum parameter $\alpha_t$ can address the blue-tilted issue before exceeding the bound imposed by BBN and CMB.

The outline of this work is as follows. We derive the gravitational wave spectrum considering a non-Bunch-Davies type vacuum as the primordial vacuum in section~\ref{sec2}. Section~\ref{sec3} discusses the connection between the pulsar timing array observation with the primordial GW spectrum generated during inflation. In section~\ref{sec4}, we constrain the inflationary and reheating parameters using NANOGrav data. We introduce the frequency dependency in $\alpha_t$ in section~\ref{sec5} to address the blue-tilted issue. Finally, we summarize our findings in section~\ref{sec6}.

\section{Gravitational wave spectrum in a non-Bunch-Davies vacuum}\label{sec2}
Primordial GW, seeded by tensor perturbations in the inflationary universe, is a plausible interpretation of the observed SGWB. Suppose the universe starts from an initial state such as a non-Bunch-Davies vacuum, which will leave its imprint on the primordial GW spectrum. In this section, we calculate the GW energy spectrum originating from a non-Bunch-Davies primordial vacuum.

In the synchronous gauge, the line element of a spatially-flat Friedmann-Lemaître-Robertson-Walker (FLRW) metric describing the inflationary tensor perturbations $h_{ij}$ reads
\begin{align}\label{LE_FLRW}
	ds^2=a^2(\tau)[-d\tau^2+(\delta_{ij}+h_{ij})dx^idx^j],
\end{align} 
where $a(\tau)$, $\tau$, $x^i$ are the scale factor, conformal time, and comoving spatial coordinates, respectively. Since the gauge-invariant tensor perturbation $h_{ij}$ is symmetric ($h_{ij}=h_{ji}$), traceless ($h_{ii}=0$), and transverse ($h_{ij,j}=0$) in nature, it contains $6-3-1=2$ independent modes. These two independent modes correspond to ``$+$" and ``$\times$" polarizations of GWs.
We treat the tensor perturbation as a quantum field in an unperturbed FLRW metric. 
Now, the second order action for the tensor perturbation $h_{ij}$ in the absence of anisotropic stress is given by~\cite{Baumann:2009ds} 
\begin{align}\label{action_TP}
	S=\int d\tau d\textbf{x}\,a^2 \frac{M_P^2}{8}\bigg[(h_{ij}')^2-(\partial_l h_{ij})^2\bigg],
\end{align} 
where $M_P^{-2}\equiv8\pi G_N$ and $G_N$ denotes the Newton's constant. Here, the prime ($~^\prime~$) denotes derivative w.r.t the conformal time. 
In Fourier space, the tensor perturbation $h_{ij}$ can be expanded as
\begin{align}\label{Fourier}
	h_{ij}(\tau,\textbf{x})=\sum_{r=+,\times}\int\frac{d\textbf{k}}{(2\pi)^{3/2}}\epsilon_{ij}^r(\textbf{{k}})h_{\textbf{k}}^r(\tau)e^{i\textbf{k}\cdot\textbf{x}},
\end{align}
where the superscript $r$ labels the modes of polarization $r$ = +/×, and $\epsilon_{ij}^r(\textbf{{k}})$ is the polarization tensor, which satisfies $\epsilon_{ij}^r(\textbf{{k}})$ $\epsilon^{ij,s}(\textbf{{k}})$ = $\delta^{rs}$.
We introduce the canonically normalized field $v^r_{\textbf{k}}\equiv\frac{a}{2}M_P h_{\textbf{k}}^r$ and utilize Eq.~(\ref{action_TP}) to obtain the second order action in the Fourier space~\cite{Mukhanov:1990me}
\begin{align}\label{Fourier_Action}
	S=\sum_{r=+,\times}\int d\tau d\textbf{k}\,\frac{1}{2}\bigg[{(v_{\textbf{k}}^r}')^2-\left(k^2-\frac{a''}{a}\right)(v_{\textbf{k}}^r)^2\bigg].
\end{align}
Now, one can canonically quantize the action by promoting the field $v_{\textbf{k}}^r(\tau)$ to operators
\begin{align}\label{BD Operator}
	\hat{v}_{\textbf{k}}^r(\tau)=v_k({\tau})\hat{a}_\textbf{k}^r+v_k^*(\tau)\hat{a}_{-\textbf{k}}^{r^{\dagger}},  
\end{align}
where the creation and annihilation operators satisfy the following canonical commutation relations
\begin{align}\label{Commutation2}
	\left[\hat{a}_\textbf{k}^r,\hat{a}_{\textbf{k}^\prime}^{s^{\dagger}}\right]
	= \delta^{rs}\delta^{(3)}(\textbf{k}-\textbf{k}^\prime), ~~ \left[\hat{a}_\textbf{k}^r,\hat{a}_{\textbf{k}^\prime}^{s}\right]
	= \left[\hat{a}_\textbf{k}^{r{\dagger}},\hat{a}_{\textbf{k}^\prime}^{s^{\dagger}}\right]=0.
\end{align}
The mode functions $v_k({\tau})$ satisfy the following equation of motion
\begin{align}\label{EOM_MS}
	v_k''+\left(k^2-\frac{a''}{a}\right)v_k=0,
\end{align}
where $k(=|\textbf{k}|$) represents the wave number. Note that, due to the isotropy of the universe, the mode functions solely depend on the time and wave number and remain independent of the direction $(\hat{\textbf{k}})$ and the polarization. In the subhorizon limit ($|k \tau| \gg 1$), one can neglect the term $a''/a$ compared to $k^2$ in Eq.~(\ref{EOM_MS}) and so the general solution of  $v_k(\tau )$ takes up the following form
\begin{align}\label{Gen. Sol}
	v_k(\tau)=F_1(k)e^{-ik\tau}+F_2(k)e^{ik\tau}.
\end{align}
In the asymptotic past ($|k \tau| \gg 1$), spacetime locally resembles Minkowski spacetime. Deep inside the horizon, the perturbation modes do not feel the effect of curvature. Therefore, we impose the initial condition that the mode functions approach the positive-frequency Minkowski solution in the subhorizon limit. The choice of mode function determines the choice of vacuum state $|0\rangle_{BD}$ for which $\hat{a}_\textbf{k}^r~|0\rangle_{BD} =0$. This standard choice, known as the Bunch-Davies vacuum~\cite{Bunch:1978yq}, corresponds to the Minkowski vacuum of a comoving observer in the distant past. The corresponding initial condition on the mode function is
\begin{align}\label{Initial}
	\lim_{|k \tau| \gg 1} v_k(\tau) = \frac{1}{\sqrt{2k}}e^{-ik\tau}.
\end{align}
Following the Bunch-Davies initial condition in Eq.~(\ref{Initial}), one can obtain the Bunch-Davies mode function 
\begin{align}\label{BD}
	h_k(\tau)=\frac{2}{M_P}\frac{e^{-ik\tau}}{a(\tau)\sqrt{2k}}\bigg(1-\frac{i}{k\tau}\bigg).
\end{align}
However, there is no unique vacuum state in the de-Sitter spacetime due to the absence of any global timelike Killing vector~\cite{Birrell:1982ix} that uniquely specifies the positive frequency mode. In general, one may choose a  different set of mode function $\tilde{h}_k(\tau)$ and correspondingly a different set of creation (annihilation) operators $\hat{b}_\textbf{k}^\dagger(\hat{b}_\textbf{k})$
\begin{align}\label{NBD Operator}	\hat{h}_{\textbf{k}}^r(\tau)=\tilde{h}_k({\tau})\hat{b}_\textbf{k}^r+\tilde{h}_k^*(\tau)\hat{b}_{-\textbf{k}}^{r^{\dagger}},
\end{align}
where two different sets of mode functions are related by Bogoliubov transformation~\cite{Allen:1985ux}
\begin{align}\label{NBD}
	{\tilde{h}}_k({\tau})&=Ah_k(\tau)+Bh^*_k(\tau) \nonumber\\
	&=\frac{2}{M_P}\frac{A e^{-ik\tau}}{a(\tau)\sqrt{2k}}\bigg(1-\frac{i}{k\tau}\bigg)+\frac{2}{M_P}\frac{Be^{ik\tau}}{a(\tau)\sqrt{2k}}\bigg(1+\frac{i}{k\tau}\bigg).
\end{align}
Here, $A$ and $B$ are the Bogoliubov coefficients which satisfy the orthonormalization condition $ |A|^2-|B|^2=1$. The operator $\hat{b}_k^r$ annihilates non-Bunch-Davies vacuum state $\hat{b}_\textbf{k}^r~|0\rangle_{NBD} =0$. Note that, unlike the Minkowski mode function, the mode function in Eq.~(\ref{NBD}) includes both positive and negative-frequency modes.

Generally, the GW background in the early universe is characterized by the tensor power spectrum, $\mathcal{P}_T(k,\tau)$, which is defined as
\begin{align}\label{tensor_PS}
	\mathcal{P}_T(k,\tau)\equiv \frac{d\langle0|\hat{h}_{ij}^2(\tau,\textbf{x})|0\rangle}{d\text{ln}k}=64\pi G_N\frac{k^3}{2\pi^2}|h_k(\tau)|^2.
\end{align}
Now, one important measurable quantity in the GW experiments is the  dimensionless GW energy spectrum $\Omega_{GW}(k)$
\begin{align}\label{def_GW}
	\Omega_{GW}(k,\tau)\equiv \frac{1}{\rho_{crit}}\frac{d\langle 0|\hat{\rho}_{GW}(\tau)|0\rangle}{d\text{ln}k},
\end{align}
where $\rho_{crit}(=3 H^2(\tau)/8\pi G_N)$ refers to critical density of the universe. 
Using the non-Bunch-Davies mode function $\tilde{h}_k(\tau)$ in Eq.~(\ref{NBD}), one can relate the GW energy spectrum to the tensor power spectrum. It is important to note that the mode function's dependence on the parameters $A$ and $B$ reveals its non-uniqueness. Different choices of these coefficients correspond to different non-Bunch–Davies initial states, including the Danielsson prescription~\cite{PhysRevD.66.023511}, the coherent state~\cite{Kundu:2011sg,Kundu:2013gha}, the Mottola-Allen vacuum~\cite{Mottola:1984ar,Allen:1985ux}, and the thermal state~\cite{Bhattacharya:2005wn}. 

Following  Ref.~\cite{Allen:1985ux}, we consider the Mottola-Allen vacuum for which the Bogoliubov coefficients are parametrized by two parameters such as $A=\cosh(\alpha)$  and $B=e^{i\beta}\sinh(\alpha)$ with $0\le\alpha<\infty$ and $-\pi<\beta<\pi$.
Note that the Bunch-Davies vacuum is a special case ($\alpha = 0$) of this kind of vacuum. We use the parametrization $\{\alpha_t,\beta_t\}$ and $\{\alpha_s,\beta_s\}$ for tensor and scalar perturbations, respectively throughout the analysis. The GW energy spectrum in the case of this non–Bunch-Davies vacuum can be expressed as
\begin{align}\label{GW_alpha}
	\Omega^{NBD}_{GW}(k,\tau)=\frac{1}{12}\frac{k^2}{a^2(\tau)H^2(\tau)}\mathcal{P}_T^{NBD}(k,\tau) ~.
\end{align}

Till now, we have evaluated the primordial quantities during inflation. However, various experiments are operating at different fixed pivot scales. For instance, the Cosmic microwave background experiment like Planck set this pivot scale $k_*=0.05~\text{Mpc}^{-1}$, whereas the GW experiments, such as PTA and laser interferometer (LI) measurements, provide their results in terms of the GW energy spectrum measured today (i.e., $\tau=\tau_0$). Therefore, it is necessary to understand how the primordial quantities evolve throughout the universe's history. In Eq.~(\ref{tensor_PS}), the tensor power spectrum defined at some conformal time $\tau_i$, shortly after the end of inflation when all the modes of interest have already left the horizon but have not yet re-entered, is related to the tensor power spectrum at a later time $\tau$ by a multiplicative transfer function $T_T(k,\tau)$, such that
\begin{align}\label{transfer_fn}
	\mathcal{P}_T(k,\tau)=T_T(k,\tau)	\mathcal{P}^{\text{prim}}_T(k,\tau_i).
\end{align}

It is important to emphasize that the primordial tensor power spectrum can be described by a power-law
\begin{align}\label{TPS_PL}
	\mathcal{P}^{\text{prim}}_T(k,\tau_i)=A_T\bigg(\frac{k}{k_*}\bigg)^{n_t}=rA_s\bigg(\frac{k}{k_*}\bigg)^{n_t},
\end{align}
where the quantities $A_T$ and $A_s$ represent the amplitudes of the primordial tensor and scalar power spectra, respectively, measured at a pivot scale $k_*$. In the context of single-field slow-roll inflation with a non-Bunch-Davies inflationary vacuum, the scalar and tensor amplitudes can be related to the slow-roll parameters and inflationary vacuum parameters as~\cite{Ashoorioon:2013eia}
\begin{align}
	&A_T^{NBD}=\frac{2}{\pi^2}\frac{H_*^2}{M_{P}^2}\times2~(\cosh^2\alpha_t+\sinh^2\alpha_t-\sinh2\alpha_t\cos\beta_t)\label{ATNBD},\\
	&A_s^{NBD}=\frac{1}{8\pi^2\epsilon_{*}}\frac{H_*^2}{M_{P}^2}(\cosh^2\alpha_s+\sinh^2\alpha_s-\sinh2\alpha_s\cos\beta_s),\label{ASNBD}
\end{align}
where $H_*$ and $\epsilon_*$ represent the Hubble rate and the first slow-roll parameter when the mode crosses the horizon during inflation, respectively. Here, the additional factor of 2 in Eq.~(\ref{ATNBD}) appears because we adopt the same parametrization for both polarizations of tensor perturbations. Thus, in the case of a non-Bunch-Davies vacuum, the tensor power spectrum $P_T^{NBD}(k,\tau)$ as well as GW energy spectrum depends explicitly on the inflationary vacuum parameters $\alpha_t\, \text{and}\, \beta_t$.

Now, the present-day GW spectrum is related to the primordial tensor spectrum as
\begin{align}\label{GW_alpha_present}
	\Omega_{GW}^0(f)= \frac{1}{12}\bigg[\frac{2\pi f}{H_0}\bigg]^2T_T(f)	\mathcal{P}^{\text{prim}}_T(f),
\end{align}
where $f=\frac{9.72\times10^{-15}}{2\pi a_0}\frac{k}{\text{Mpc}^{-1}}$~Hz is the present-day physical frequency of the GW associated with the comoving wavenumber $k$ and $H_0$ refers to the present-day Hubble rate. The tensor transfer function can be derived using the Ref.~\cite{Boyle:2005se} as 
\begin{equation} \label{transfer_expr}
	T_T(f) =
	\begin{dcases}
		\frac{0.4\times C_2(\omega_{\text{re}})}{(1+z_{\text{re}})^2}\bigg[\frac{\tilde\gamma^{-1/2}2\pi f}{(1+z_{\text{re}})H_0}\bigg]^\frac{-4}{1+3\omega_{\text{re}}}\hspace{2.3cm} \text{for}~~f\ge \frac{k_{re}}{2\pi},\\
		3.7\times10^{-5}\times\bigg[\frac{2\pi f}{H_0}\bigg]^{-2} \hspace{4.1cm} 
		\text{for}~~\frac{k_{\text{eq}}}{2\pi}\le f\le \frac{k_{re}}{2\pi},\\
		0.22\times\bigg[\frac{2\pi f}{H_0}\bigg]^{-4} \hspace{5.22cm} \text{for}~~f\le\frac{k_{\text{eq}}}{2\pi},
	\end{dcases}
\end{equation}
where $k_{re}(=1.7\times10^{14}(T_{\text{re}}/10^7~\text{GeV})~\text{Mpc}^{-1})$ and $k_{\text{eq}}(=1.01\times10^{-2}~\text{Mpc}^{-1})$ correspond to the wavenumbers at the end of reheating and matter-radiation equality, respectively. The factor $\tilde \gamma \,(=8.03\times10^{-5})$ is a constant number and $C_2({\omega_\text{re}})$ is given by~\cite{Boyle:2005se}
\begin{align}\label{c2f}
	C_2({\omega_\text{re}})= \frac{\bigg[\Gamma\left(\frac{5+3\omega_{\text{re}}}{2(1+3\omega_{\text{re}})}\right)\bigg]^2}{\pi}\; \left[1+3\omega_{\text{re}}\right]^\frac{4}{1+3\omega_{\text{re}}},
\end{align}
where $\Gamma(..)$ is the Gamma function and the redshift at the end of reheating ($z_{\text{re}}$) is connected with the reheating temperature as 
\begin{align}\label{z_re}
	1+z_{\text{re}}=\frac{1+z_{\text{eq}}}{T_{\text{eq}}}\,T_{\text{re}}.
\end{align}
The wavenumber at the end of reheating $k_{\text{re}}$ is related to the reheating temperature through the horizon crossing condition i.e., $k_{\text{re}}=a_{\text{re}}H_{\text{re}}$ where $a_{\text{re}}$ $(H_{\text{re}})$ refers to the corresponding scale factor (Hubble rate). Using the entropy conservation and the expression of the Hubble parameter at the end of reheating $H_{\text{re}}\left(=\frac{\pi}{3}\sqrt{\frac{g_{*\text{re}}}{10}}\frac{T_{\text{re}}^2}{M_P}\right)$ with $g_{*\text{re}}(=106.75)$ being the effective number of relativistic degrees of freedom, we obtain
\begin{align}\label{k_re}\nonumber
	k_{\text{re}}&=a_0\frac{T_0}{T_{\text{re}}}\bigg(\frac{g_{*s,0}}{g_{*s,\text{re}}}\bigg)^{\frac{1}{3}}\frac{\pi}{3}\sqrt{\frac{g_{*\text{re}}}{10}}\frac{T_{\text{re}}^2}{M_P}\\
	&=1.7\times10^{14}\left(\frac{T_{\text{re}}}{10^7~ \text{GeV}}\right) \text{Mpc}^{-1}
\end{align}
where subscript $0$ denotes present-day quantities and $g_{*s}$ refers to the entropic degrees of freedom. 

\section{Pulsar timing arrays: Link with the physics of the early universe}\label{sec3}
The GW spectral energy density associated with the PTA experiments is typically expressed as
\begin{align}\label{omega_PTA}
	\Omega_{GW}^{\text{PTA}}=\frac{2\pi^2}{3H_0^2}f^2h_c^2(f),
\end{align}
where $h_c(f)$ refers to the power spectrum of GW strain, measured at the reference frequency, $f_{\text{yr}}=1\, \mathrm{yr}^{-1}\approx3.17\times10^{-8}$ Hz. Conventionally $h_c(f)$ is parametrized by a power law with amplitude $A$ and spectral index $\delta$ as
\begin{align}\label{GW_strain}
	h_c(f)=A\bigg(\frac{f}{f_{\text{yr}}}\bigg)^\delta,
\end{align}
where $\delta$ is further related to the pulsar timing residual cross-power density index $\gamma$ $(\delta=(3-\gamma)/2)$. It is noteworthy to mention that the NANOGrav 15-year dataset, which measured the characteristic strain in the frequency range $f \in [2\times10^{-9},6\times10^{-8}]$, is modelled as a power law using the form of the GW energy density
\begin{align}\label{omega_fitted}
	\Omega_{GW}^{\text{PTA}}=A^2\frac{2\pi^2}{3H_0^2}\frac{f^{5-\gamma}}{f_{\text{yr}}^{3-\gamma}}.
\end{align}
PTA data from NANOGrav constrains $A$ and $\gamma$ as a joint posterior distribution in addition to inferring the best fit values of $A=6.4^{+4.2}_{-2.7}\times10^{-15}$ and $\gamma=3.2\pm0.6$, at a reference frequency $1 \,\text{yr}^{-1}$~\cite{NANOGrav:2023gor}. Now, it is important to connect the early universe parameters with the quantities $A$ and $\gamma$. The relation among the parameters $\gamma$, $n_t$, and $\omega_{\text{re}}$ can be obtained by equating Eq.~(\ref{GW_alpha}) with Eq.~(\ref{omega_fitted})
\begin{align}\label{resudial_index}
	\gamma=5-n_t+2\zeta,
\end{align}
with $\zeta=\dfrac{1 - 3\, \omega_{\text{re}}}{1 + 3\,\omega_{\text{re}}}$ and the amplitude of the PTA signal is guided by
\begin{align}\label{strain_amp}
	A=\bigg(\frac{rA_sC_2 H_0^2}{20\pi^2}\bigg)^{1/2}\tilde{\gamma}^{(1+\zeta)/2}(2\pi)^{(n_t-2\zeta)/2}\bigg(\frac{a_0}{k_*}\bigg)^{n_t/2}\bigg(\frac{T_{\text{eq}}}{H_0(1+z_{\text{eq}} )T_{\text{re}}}\bigg)^{-\zeta}{f^{(1+\zeta-n_t/2)}_\text{yr\,\,\,}}.
\end{align}
Here, we consider that all the frequencies within the frequency range for which NANOGrav measures the characteristic strain, re-enter the horizon during the reheating phase, thus the amplitude $A$ becomes a function of $n_t$, $r$, and $T_{\text{re}}$. We have taken the following values for the cosmological parameters $k_{\text{eq}}= 1.01 \times 10^{-2}~\text{Mpc}^{-1}$, $z_{\text{eq}}=3400$, $T_{\text{eq}}\simeq  1.25\times 10^{29}~\text{Mpc}^{-1}$, $\tilde{\gamma}\simeq 8.03\times 10^{-5}$, and $	\mathcal{P}^{\text{prim}}_T(k_*,\tau_i)(\equiv A_s)=2.1\times 10^{-9}$ at $k_*=0.05~\text{Mpc}^{-1}$ from Planck 2018 results~\cite{Planck:2018vyg}. Being massless degrees of freedom, the GW contributes to the radiation energy density of the early universe. This additional contribution is quantified by $\Delta N_{\text{eff}}$, which is restricted by BBN as well as CMB. The SGWB contribution to $\Delta N_{\text{eff}}$ is characterized by~\cite{Kuroyanagi:2014nba,Ben-Dayan:2019gll,Vagnozzi:2022qmc,Giare:2022wxq}
\begin{align}\label{N_eff}
	\int_{f_{\text{min}}}^{f_{\text{max}}} df\; \frac{\Omega_{GW}h^2}{f}\simeq5.6\times10^{-6}\Delta N_{\text{eff}},
\end{align}
where ${f_{\text{min}}}$ and ${f_{\text{max}}}$ are the frequencies that depend on the epoch of interest and the epoch where the temperature of the universe reached its maximum after the Big Bang, respectively. We will come back to this in section~\ref{sec4} to elucidate how constraints on $\Delta N_{\text{eff}}$ place an upper bound on the GW relic density.

\section{Constraints on inflationary and reheating parameters}\label{sec4}
The NANOGrav collaboration provides strong evidence for a stochastic signal with a common amplitude and spectrum by analysing their latest 15 years of data. In this section, we constrain the parameters of the early universe utilizing this dataset by performing a Markov chain Monte Carlo (MCMC) analysis. In Ref.~\cite{Vagnozzi:2023lwo}, the inferred NANOGrav 15-year constraints on $A$ and $\gamma$ are translated to restrict the inflationary parameters ($n_t,~r$), and we closely follow this analysis. It is noteworthy that we further generalize the analysis by constraining the reheating era in addition to restricting the inflationary parameters. Interestingly, NANOGrav $A-\gamma$ joint posterior distribution is well approximated by a bivariate Gaussian in $\log_{10}A$ and $\gamma$, with mean vector $\mu_{15}$ and covariance matrix $\Sigma_{15}$ given by~\cite{Vagnozzi:2023lwo}
\begin{equation}\label{mean}\nonumber
	\mu_{15}\approx(-14.20, 3.20),
\end{equation}
\begin{equation}\label{cov mat}\hspace{0.9cm}
	\Sigma_{15}\approx \begin{pmatrix}
		0.127 & -0.045 \\
		-0.045 & 0.021
	\end{pmatrix}.
\end{equation}
We define the log-likelihood function for NANOGrav 15 years using the mean vector $\mu_{15}$ and $\Sigma_{15}$ as follows
\begin{align}\label{likelihood}
	\ln\mathcal{L}(\theta)=-\frac{(x(\theta)-\mu_{15})^T\Sigma_{15}^{-1}(x(\theta)-\mu_{15})}{2},
\end{align}
where $x(\theta)$ refers to the derived vector parameters $x(\theta)\in\{\log_{10}A(\theta),\gamma(\theta)\}$ with $\theta$ stands for the vector of the inflationary and reheating parameters. In general, $\theta$ should contain all the relevant cosmological parameters as well as reheating parameters describing the shape and amplitude of the inflationary SGWB: $r,n_t,\Omega_{m0}, A_s, H_0, T_{\text{re}},\omega_{\text{re}}$. Although we mentioned earlier that we have taken the best-fit values for all the cosmological parameters except $r$ and $n_t$, as cosmological observations, particularly the Planck satellite, measured these parameters with great precision. Here, we assume $\theta_1\equiv\{r,n_t,\omega_{\text{re}}\}$ for some fixed values of reheating temperature.

Now, we proceed with the vector $\theta_1$ and relate $r$, $n_t$ and $\omega_{\text{re}}$ to the derived parameters $\log_{10}A$ and $\gamma$ entering the NANOGrav likelihood of Eq.~(\ref{likelihood}) via Eqs.~(\ref{resudial_index}) and (\ref{strain_amp}). By adding the NANOGrav log-likelihood to the \texttt{cobaya}~\cite{Torrado:2020dgo} package, we generate the MCMC chain and set a convergence criterion $R-1<0.001$. We work with $\log_{10}r$ instead of $r$, so our actual $\theta_1$ is $\{\log_{10}r, n_t, \omega_{\text{re}}\}$. In our analysis, we consider a flat prior on $\textcolor{blue}{\log_{10}r\in[-250,-1.44]}$, $n_t\in[0,10]$, and $\omega_{\text{re}}\in[-1/3,1]$, while fixing the reheating temperature to 50 MeV. Additionally, we impose the $2\sigma$ upper limit on $\log_{10}r$, placed by the joint analysis of Planck, WMAP, BICEP2, and BICEP3 data~\cite{BICEP:2021xfz}.
\begin{figure}[htb!]
	\centering
	\includegraphics[width=12cm,height=12cm]{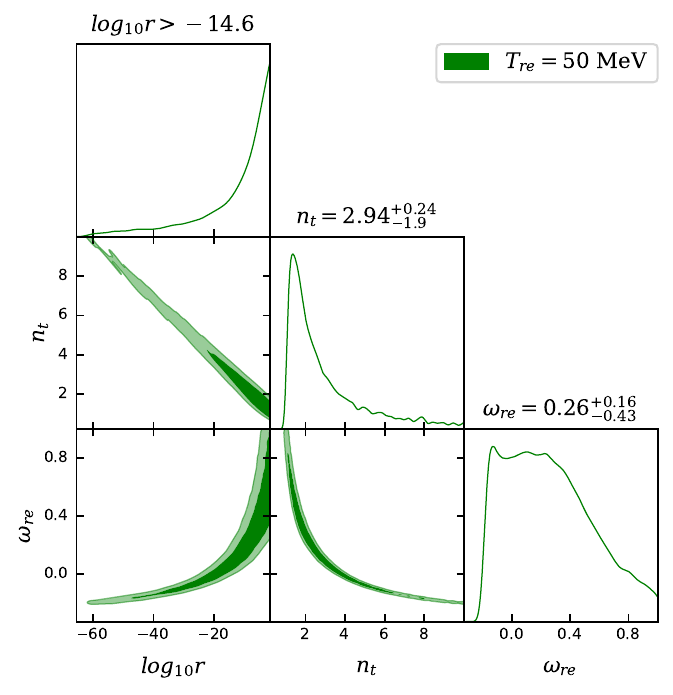} 
	\caption{Posterior distributions of inflationary and reheating parameters ($\log_{10}r, n_t, \omega_{\text{re}}$) for fixed reheating temperature, $T_{\text{re}}=50$ MeV.}
	\label{triangular}
\end{figure}
The triangular plot of Figure~\ref{triangular} depicts the posterior distributions for $r$, $n_t$, and $\omega_{\text{re}}$, where the dark green shaded area corresponds to $68\%$ CL probability region and the light green shaded area represents $95\%$ CL probability region. Our fits show that $n_t=2.94^{+0.24}_{-1.9}$, $\omega_{\text{re}}=0.26^{+0.16}_{-0.43}$, and the tensor-to-scalar ratio is bounded from below, $r>2.51\times10^{-15}$. Therefore, NANOGrav favours an extremely blue-tilted tensor spectrum, while both matter-like and radiation-like reheating scenarios remain consistent with the observations. As the Planck data only sets an upper limit on the tensor-to-scalar ratio, in this work, we adopt a lower bound of $\log_{10} r = -250$. Additionally, we have verified that our results remain unchanged even when smaller values of $\log_{10} r$ are considered.

To constrain the reheating phase of the universe using NANOGrav, one needs to ensure that the primordial GW tensor modes must re-enter the horizon during the reheating phase of the universe. Given the frequency range of the NANOGrav data, the permissible reheating temperature range thus becomes $T_{\rm re} \approx 4-50$ MeV, where the lower bound on $T_{\rm re}$ is set by the BBN constraints. Ref.~\cite{Ben-Dayan:2023lwd} has also attempted to constrain the reheating phase of the early universe, considering an instantaneous reheating scenario. This analysis has not considered the aforementioned criterion, but rather considers $T_{\text{re}}$ as a free parameter, taking a much wider range, in particular $T_{\text{re}}\in[10^{-5}$ GeV-$10^{15}~\text{GeV}]$. In our work, we focus on a non-instantaneous reheating scenario while accounting for the necessary upper limit on $T_{\text{re}}$.
Although we fixed the reheating temperature in Figure~\ref{triangular}, we have checked that our results remain largely unchanged if the reheating temperature varies within the permitted range of 4-50 MeV. It is clear from our analysis that the non-instantaneous reheating scenario is favoured by the NANOGrav observations.

Interestingly, the same framework can be extended to constrain the parameters characterizing the inflationary vacuum. In particular, the amplitude of inflationary SGWB depends on the combination $rA_s \equiv A_T$. Instead of adopting the best-fit value of the scalar amplitude $A_s$ from the PLANCK experiment, we can employ the expression in Eq.~(\ref{ATNBD}) for the tensor amplitude corresponding to the non-Bunch-Davies vacuum, considering a single-field inflation scenario.
On the CMB pivot scale ($k_*=0.05~\text{Mpc}^{-1}$), scalar amplitude $A_s$ and the tensor-to-scalar ratio $r$ are restricted to be small by the Planck data. Note that the smallness of $A_T$ ($\equiv rA_s$) necessitates maximum cancellation among the terms within the bracket in Eq.~(\ref{ATNBD}), which leads to the requirement that $\beta_t$ must be equal to $0$. Therefore, we can say that observational evidence favours an alpha-vacuum, which is a specific choice within the adopted parametrization of non-Bunch-Davies vacua. It is important to emphasize that alpha vacuum is known to possess several theoretical shortcomings\footnote{Alpha-vacuum states are non-Hadamard states, exhibiting additional antipodal singularities in their two-point functions, which lead to non-renormalizable ultraviolet divergences in interacting quantum field theory on de Sitter spacetime~\cite{Banks:2002nv,Einhorn:2002nu,Collins:2003mk,Collins:2003mj}. These additional singularities give rise to non-local divergences, causing the standard local renormalization procedure to fail. }. However, these theoretical issues do not affect the phenomenological analysis performed here as we restrict our analysis to the tree-level tensor power spectrum and the resulting stochastic GW spectrum. Instead, the alpha-vacuum should be treated as
an effective parametrization of non-standard initial conditions, while its complete theoretical
consistency remains beyond the scope of the present work. 

With the choice $\beta_t=0$, we can simplify the expression for $A_T$ as follows: 
\begin{align}\label{A_T_new}
	A_T^{NBD}=&\frac{4}{\pi^2}\frac{H_*^2}{M_{P}^2}e^{-2\alpha_t}.
\end{align}
This allows us to define an alternative vector $\theta_2\equiv\{\alpha_t,n_t,\omega_{\text{re}}\}$ for some fixed values of reheating temperature and the ratio ${H_*}/{M_{P}}$.

We then repeat the MCMC analysis using the vector $\theta_2$, mapping $\alpha_t$, $n_t$ and $\omega_{\text{re}}$ to the parameters $\log_{10}A$ and $\gamma$ of the NANOGrav log-likelihood function in Eq.~(\ref{likelihood}) via Eqs.~(\ref{resudial_index}), (\ref{strain_amp}), and (\ref{A_T_new}). In this case, we consider flat prior on $\alpha_t\in[0.38,5000]$, $n_t\in[0,10]$, and $\omega_{\text{re}}\in[-1/3,1]$ while fixing reheating temperature to 50 MeV and the ratio ${H_*}/{M_{P}}$ to $2\times10^{-5}$. This choice of ${H_*}/{M_{P}}$ is consistent with the upper limit on the tensor-to-scalar ratio inferred from Planck observations. The lower limit on $\alpha_t$, i.e., $0.38$ is chosen such that it is consistent with the upper bound on the tensor-to-scalar ratio $r(\equiv{A_T}/{A_s})$ obtained from the Planck experiment. In addition, we have also used the best-fit value of $A_s(=2.1\times10^{-9})$ obtained from the Planck experiment. Although the theoretical bound on the parameter $\alpha$ extends to $\infty$, we restrict our analysis to $\alpha_t <5000.$ We have verified that increasing the upper bound beyond this value does not affect our results.
\begin{figure}[htb!]
	\centering
	\includegraphics[width=12cm,height=12cm]{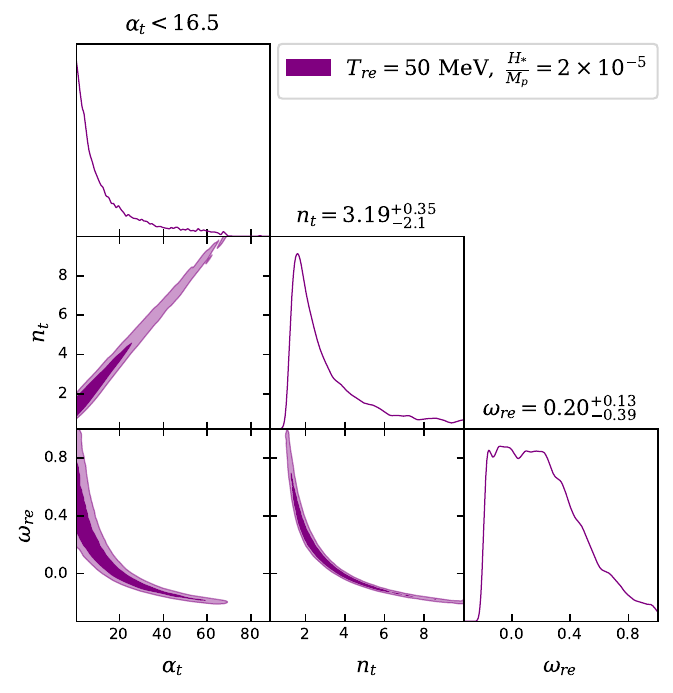} 
	\caption{Posterior distributions of inflationary and reheating parameters ($\alpha_t, n_t, \omega_{\text{re}}$) for fixed reheating temperature $T_{\text{re}}=50$ MeV and $\frac{H_*}{M_{P}}=2\times10^{-5}$.}
	\label{triangular2}
\end{figure}

The triangular plot of Figure~\ref{triangular2} presents the posterior distributions for $\alpha_t$, $n_t$, and $\omega_{\text{re}}$, where the dark purple shaded area corresponds to $68\%$ CL probability region and the light purple shaded area represents $95\%$ CL probability region. The inferred values of $n_t$ and $\omega_{\text{re}}$ remains consistent with the previous case $n_t=3.19^{+0.35}_{-2.1}$ and $\omega_{\text{re}}=0.20^{+0.13}_{-0.39}$. In addition, the inflationary vacuum parameter is constrained from above, yielding $\alpha_{t}<16.5$. This value is consistent with the bound~\cite{Lyth:1998xn,Chung:1999ve,Cielo:2024poz} required to ensure that any potential backreaction effects do not affect inflationary dynamics. In deriving this bound, we fix the ratio ${H_*}/{M_{P}}$ to $2\times 10^{-5} $ corresponding to the current upper bound on the tensor-to-scalar ratio from Planck data. Choosing a smaller value of ${H_*}/{M_{P}}$ is equally consistent with the observational bounds. We have explicitly verified that reducing $H_*/M_P$ leads to a tighter upper bound on $\alpha_t$, thereby providing an even stronger suppression of possible backreaction effects. A brief discussion of the backreaction constraint can be found in the appendix~\ref{backreaction}.

In summary, NANOGrav 15-year data place strong constraints on the tensor spectral index and the reheating equation of state. GWs contribute to the effective number of neutrino species ($N_{\text{eff}}$) since it act as an additional component of radiation. As a result, any contribution from the GW is constrained by CMB and BBN observations. Using the integral in Eq.~(\ref{N_eff}), we evaluate the contribution of PTA signal to $N_{\text{eff}}$. We obtain an upper limit on the GW relic density by using the upper bound on $\Delta N_{\text{eff}}$ from Planck data. This data $(\Delta N_{\text{eff}}\le 0.04$~\cite{Planck:2018vyg}) dictates that $\Omega_{GW}h^2$ larger than $2.4\times10^{-6}$ is excluded by BBN~\cite{Aver:2015iza, Cooke:2017cwo, Mossa:2020gjc}. The large positive value of the spectral index $n_{t}$ yields a steeper SGWB spectrum, preferred by NANOGrav, that violates BBN constraints and gives rise to the so-called blue-tilted issue.

\section{Blue-tilted spectrum and frequency dependent alpha-vacuum}\label{sec5}

A blue-tilted GW energy spectrum, when extrapolated to high frequencies, can violate BBN constraints. In this section, we demonstrate that introducing a frequency-dependent parametrization of the vacuum parameter $\alpha_t$ provides a viable resolution to this issue.

\begin{figure}[htb!]
	\centering
	\includegraphics[width=12cm,height=8cm]{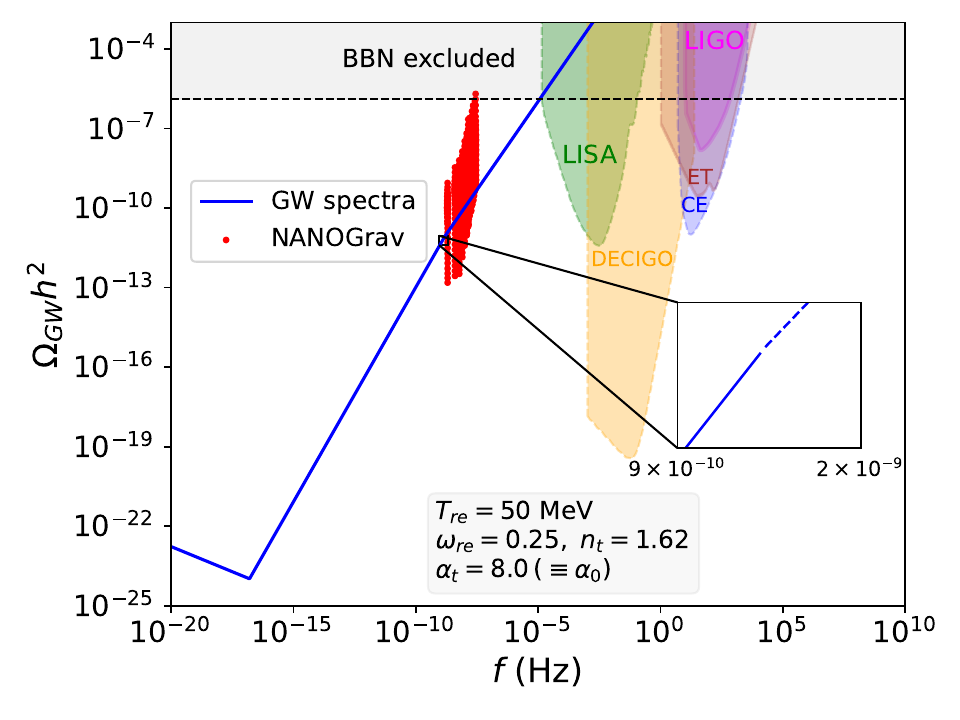} 
	\caption{GW relic density as a function of frequency for a fixed $\omega_{\text{re}}$, $n_t$, and $\alpha_t$ which are within the $95\%$ CL region constrained by NANOGrav. The gray-shaded region is ruled out by BBN, and the red dots depict the NANOGrav 15-year data. Coloured regions with dotted contours indicate the projected sensitivity band of future experiments, while coloured regions with solid contours denote current bounds from ongoing experiments. The green, orange, blue, and brown shaded regions correspond to anticipated sensitivity curves of upcoming GW detectors, namely LISA, DECIGO, CE, and ET, respectively. The magenta-shaded region indicates the area excluded by the LIGO experiment.}
	\label{GW_spectra}
\end{figure}

Before introducing a frequency-dependent parametrization, we first study the behaviour of the GW energy spectrum as a function of frequency. For illustration, we choose a benchmark from the $95\%$ CL contour of $n_t$ and $\omega_{\text{re}}$ in Figure~\ref{triangular2} to depict the GW spectrum. Figure~\ref{GW_spectra} shows the GW relic density as function of frequency for $\omega_{\text{re}}=0.25$, $n_t=1.62$, $\alpha_t=8.0$ with reheating temperature fixed at $T_{\text{re}}=50$ MeV. 
Here, the gray-shaded region is excluded due to constraints coming from BBN. The red dots represent the NANOGrav 15-year data~\cite{NANOGrav:2023hvm}. The evolution of the universe can also be understood by looking at the slope of the plot in Figure~\ref{GW_spectra}, where changes in slope occur at $f \sim 10^{-17}$ Hz during the matter-radiation equality and at nHz frequency during radiation-reheating equality as highlighted in Figure~\ref{GW_spectra} inset. Given the steep slope needed to account for the NANOGrav observation, the GW spectrum enters the BBN-excluded region around $10^{-5}$ Hz soon after the NANOGrav frequency regime.

According to the joint posterior distribution of $\omega_{re}$ and $n_t$ in Figure \ref{triangular2}, increasing $\omega_{re}$ results in a smaller value of $n_t$, and vice versa. It turns out that for such a combination of $\omega_{re}$ and $n_t$ the GW spectrum will violate the BBN constraints at a smaller frequency compared to what is observed in Figure~\ref{GW_spectra}. Also, for some allowed combinations of $\omega_{re}$ and $n_t$, the GW spectrum will be so steep that it will hit the BBN constraint even before satisfying the NANOGrav data. Therefore, the parameter values in Figure \ref{GW_spectra} are chosen such that the resulting spectrum is consistent with the NANOGrav data. Also, a smaller value of $\alpha_t$ results in a larger amplitude of the GW spectrum. Accordingly, the vacuum parameter $\alpha_t$ is chosen to ensure consistency with the data. Note that, during the reheating epoch, the frequency-dependency of the GW spectrum in Eq.~(\ref{GW_alpha_present}) is given by $\Omega_{GW} \propto f^{2+n_t-\chi}$ where $\chi=4/(1+3\omega_{re})$. So, a smaller value of the reheating equation of state $\omega_{re}$ implies that the GW spectrum exceeds the BBN bound at a higher frequency. Accordingly, we set $\omega_{re}=0.25$ for which the GW spectrum hits the BBN bound at  $f_S \sim 10^{-5}$ Hz.

\begin{figure}[htb!]
	\centering
	\includegraphics[width=12cm,height=8cm]{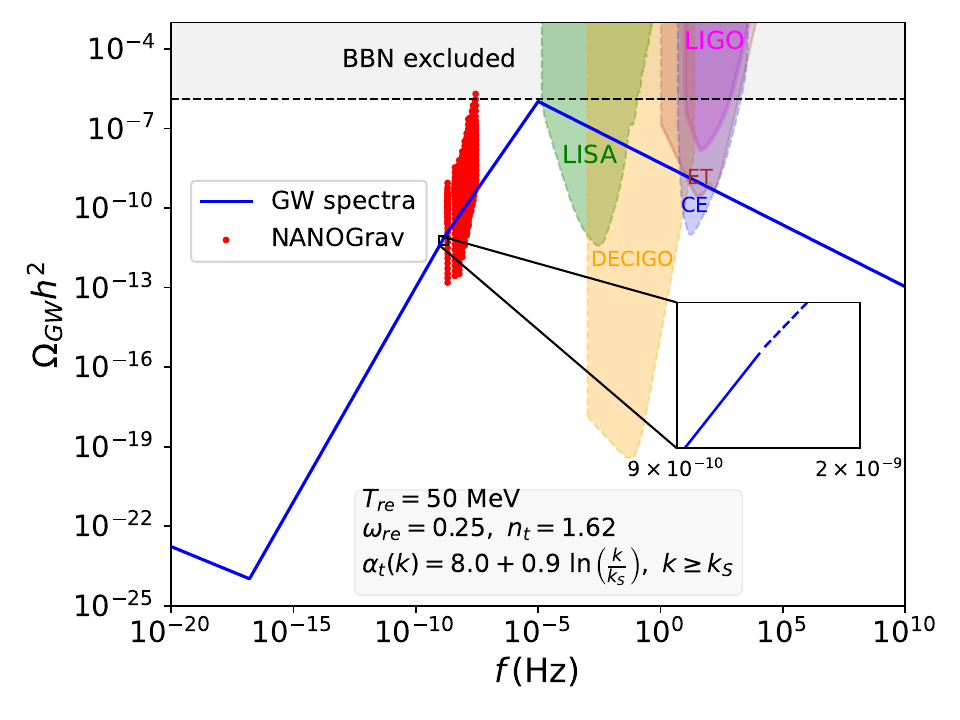} 
	\caption{Same as Figure~\ref{GW_spectra} but with frequency dependent vacuum parametrization as shown in Eq.~(\ref{alpha_cases}). The threshold frequency is $f_S=10^{-5}$ Hz.}
	\label{GW_spectra_final}
\end{figure}

An interesting approach to address the blue-tilted issue is to introduce frequency dependence in $\alpha_t$. As discussed previously in Eq.~(\ref{A_T_new}), the primordial tensor power spectrum in the case of alpha-vacuum is
\begin{align}\label{AT_new_2}
	P_T^{NBD}=&\frac{4}{\pi^2}\frac{H_*^2}{M_{P}^2}e^{-2\alpha_t}\bigg(\frac{k}{k_*}\bigg)^{n_t}.
\end{align}
The tensor spectral index is defined as $n_t = \frac{d \ln P_T}{d \ln k}$. If the vacuum parameter $\alpha_t$ is allowed to have a frequency dependence, then the tensor spectral index in the presence of a non-Bunch-Davies vacuum takes the form 
\begin{align}\label{freq_alpha}
	\frac{d \ln P_T^{NBD}}{d \ln k}=n_t-2\frac{d \alpha_t}{d \ln k}~.
\end{align}
 For a frequency-dependent vacuum parametrization, the scaling of relic density of GW modifies as $\Omega_{GW} \propto f^{2+n_t-\chi-\xi(k)}$ where $\xi(k)=2~{d \alpha_t}/{d \ln k}$. If the frequency-dependent term $\xi(k)$ exceeds the combination $2+n_t-\chi$, the GW energy spectrum remains consistent with the BBN bound. For simplicity, we may parametrize $\xi(k)=2D$ with $D>0$. However, one must be careful while considering this frequency dependence, since if it persists across the entire frequency range, the resulting GW spectrum may fail to remain consistent with NANOGrav observations. So, any frequency dependence, if present, should arise beyond a certain scale. Therefore, we adopt a frequency-dependent parametrization given by
\begin{equation} \label{alpha_cases}
	\alpha_t(k) =
	\begin{dcases}
		\alpha_0\hspace{2.88cm} \text{for}~~k\le k_S,\\
		\alpha_0+D\ln\left(\frac{k}{k_S}\right)\hspace{0.5cm}
		\text{for}~~k\ge k_S.
	\end{dcases}
\end{equation}
Here, the comoving scale $k_S$ or the corresponding frequency $f_S$ is chosen phenomenologically to remain consistent with the constraint imposed by BBN. In Figure~\ref{GW_spectra_final}, we present the GW relic density as a function of frequency, employing the frequency-dependent parametrization described above. The same parameter values as those used in Figure~\ref{GW_spectra} are adopted here. The slope of the GW spectrum after $f_S=10^{-5}$ Hz is governed by the parameter $D$. We choose $D=0.9$ in Figure~\ref{GW_spectra_final} to be consistent with the upper bound at the LIGO scale. We find that a smaller value of $D$ causes the GW spectrum to exceed the upper bound at the LIGO scale. We have checked that the backreaction effects~\cite{Lyth:1998xn,Chung:1999ve,Cielo:2024poz} appearing as a result of the frequency-dependent parametrization do not spoil the inflationary dynamics. We note that the predicted GW spectrum lies within the projected sensitivity bands of the upcoming GW experiments like LISA~\cite{LISA:2017pwj}, DECIGO~\cite{Seto:2001qf}, Cosmic Explorer (CE)~\cite{Reitze:2019iox}, Einstein Telescopes (ET)~\cite{Punturo:2010zz} etc. This allows forthcoming observations to probe the validity of the frequency-dependent parametrization.

We demonstrate that a frequency-dependent alpha-vacuum provides a solution to the long-standing blue-tilted problem when the frequency dependence is introduced beyond a comoving scale $k_S$. However, several solutions have been proposed in the literature to address the blue-tilt issue. These include incorporating astrophysical contributions to GWs \cite{Ben-Dayan:2023lwd}, running of the tensor spectral index \cite{Ben-Dayan:2023lwd}, and the suppression of the GW spectrum due to late-time entropy injection resulting from early matter domination \cite{Datta:2023xpr, Athron:2024fcj}. Notably, our study presents a minimal solution to the so-called blue-tilted problem.

Before analysing the cosmological consequences of the frequency-dependent parametrization, we examine the reheating scenarios permitted within the non-Bunch--Davies framework consistent with Figures \ref{GW_spectra} and \ref{GW_spectra_final}. From Figures \ref{triangular} and \ref{triangular2}, we note that both matter-like $(\omega_{\text{re}}=0)$ and radiation-like $(\omega_{\text{re}}=1/3)$ reheating scenarios are within the $68\%$ CL region for both Bunch-Davies and non-Bunch-Davies vacuum case. In Figure \ref{GW_spectra}, we adopt the benchmark values $\omega_{\text{re}}=0.25$ and $n_t=1.62$, for which $\alpha_t=8.0$ is required to satisfy NANOGrav observation. In contrast, the matter-like reheating scenario $(\omega_{\text{re}}=0)$ requires $n_t\sim 3.8$, for which the required value of $\alpha_t$ is around 27. This value of $\alpha_{t}$ violates the upper limit obtained in our analysis (see Figure \ref{triangular2}). Since $\omega_{\text{re}}$ and $n_t$ are negatively correlated, a smaller value of $\omega_{\text{re}}$ requires a larger value of $n_t$, which in turn demands a larger $\alpha_t$. Consequently, benchmark models with a smaller reheating equation-of-state parameter (around $\omega_{\text{re}}=0$) are excluded as they fail to address the NANOGrav observation and also violate backreaction constraints. In summary, even though both matter-like and radiation-like reheating remain viable within the Bunch-Davies vacuum framework, the NANOGrav observations and backreaction constraint effectively rule out the matter-like reheating scenario in case of the non-Bunch-Davies inflationary vacuum.

\begin{figure}[htb!]
	\centering
	\includegraphics[width=12cm,height=8cm]{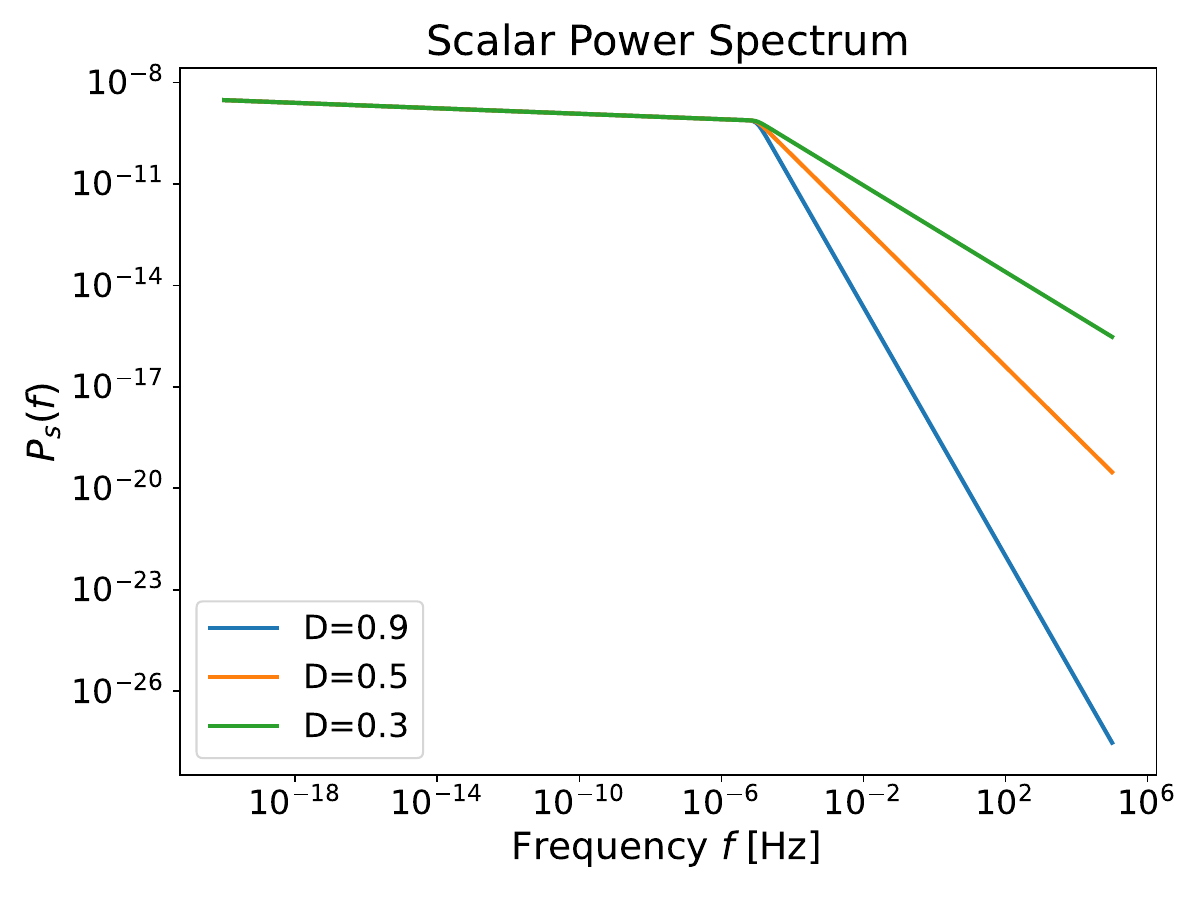} 
	\caption{Scalar power spectrum in presence of a frequency-dependent vacuum parameter $\alpha_s$.}
	\label{modified_scalar}
\end{figure}

Finally, we analyse the cosmological consequences of this frequency-dependent parametrization on the scalar sector. Assuming an analogous vacuum parametrization for scalar and tensor perturbations, the scalar amplitude given in Eq.~(\ref{ASNBD}) takes the form 
\begin{equation}\label{A_s_new}
    A_s^{NBD}=\frac{1}{8\pi^2\epsilon_{*}}\frac{H_*^2}{M_{P}^2}e^{-2\alpha_s}.
\end{equation}
On the CMB pivot scale, the best-fit value of $A_s$ from the Planck experiment is $2.1\times10^{-9}.$ Fixing $H_{*}/M_P=2\times10^{-5}$ and choosing $\alpha_s=8.0$, Eq.~(\ref{A_s_new}) yields $\epsilon_{*}\simeq 2.7\times10^{-10}$ where $*$ denotes the value of the parameter $\epsilon$ evaluated at the time of horizon crossing of the pivot scale. Having fixed the scalar amplitude at CMB scales, we now turn to the possible behaviour of the spectrum at smaller scales. As elucidated in the tensor sector, resolving the blue-tilted issue in the alpha-vacuum requires the introduction of the frequency dependency in the vacuum parameter $\alpha_t$ beyond $f_S=10^{-5}$ Hz. This transition frequency corresponds to small scales during inflation. The scalar power spectrum is known to be nearly scale-invariant ($n_s\simeq1$) in the CMB pivot scale ($k_*=0.05~\text{Mpc}^{-1}$)~\cite{Planck:2018vyg}. However, the behaviour of the scalar spectrum on smaller scales remains currently observationally unconstrained. 
It is therefore important to explore whether a frequency-dependent vacuum parametrization can modify the scalar spectrum in this regime.
Assuming that the vacuum parametrization for scalar and tensor modes is identical, i.e., $\alpha_t=\alpha_s$ and $\alpha_s$ follows the same frequency dependence as $\alpha_t$, the scalar spectral index is modified according to
\begin{align}\label{freq_alpha_Scalar}
	\frac{d \ln P_s^{NBD}}{d \ln k}=n_s-2\frac{d \alpha_s}{d \ln k}~.
\end{align}
Adopting the same parametrization for $\alpha_s$ as given in Eq.~(\ref{alpha_cases}), the scalar spectral index above the transition frequency $f_S=10^{-5}$ Hz becomes 
\begin{align}\label{freq_alpha_Scalar_2}
	\frac{d \ln P_s^{NBD}}{d \ln k}=n_s-2D
\end{align}
where $D>0$.  For consistency with the analysis of the tensor spectrum, we first consider the benchmark value $D=0.9$,  which is primarily motivated by the LIGO constraints on the GW spectrum. To illustrate the phenomenological impact of the vacuum parameter on the scalar spectrum, we also consider smaller values of $D$. The resulting modified scalar power spectra for different choices of $D$ are shown in Figure \ref{modified_scalar}.
We observe that the scalar power spectrum is nearly scale-invariant at large scales. However, for frequencies exceeding $f_S=10^{-5}$ Hz, it becomes suppressed on smaller scales, with the level of suppression governed by the parameter $D$.

So far, we have discussed solutions to the blue-titled issue, although it is important to note whether such a large $n_t~(\sim 2.94)$ value is realizable within inflationary models. In case of the standard slow-roll single field inflation model,  the spectra should be red-tilted ($n_t<0$) because of the relation, $n_t=-r/8$ with $r\ge0$~\cite{Copeland:1993ie}. Although the relation between $n_t$ and $r$ is modified in multi-field inflation~\cite{Price:2014ufa}, the spectrum remains red-tilted. Thus, the required blue-tilted spectra ($n_t\sim 2.94$) are impossible to achieve in either a single-field or a multi-field inflationary model. Interestingly, there are inflationary scenarios in which a blue-titled tensor spectrum can be realized, such as G-inflation~\cite{Kobayashi:2010cm}, elastic inflation~\cite{Gruzinov:2004ty}, super inflation~\cite{Baldi:2005gk}, solid inflation~\cite{Endlich:2012pz, Akhshik:2014gja}, beyond slow-roll inflation~\cite{Gong:2014qga}, etc. Additionally, non-inflationary scenarios like string gas cosmology~\cite{Brandenberger:1988aj,Nayeri:2005ck,Brandenberger:2006xi,Brandenberger:2006vv,Brandenberger:2014faa}, ekpyrotic scenario~\cite{Khoury:2001wf,Brandenberger:2020tcr}, bouncing cosmology~\cite{Brandenberger:2009jq}, etc. can also lead to the generation of a blue-tilted GW spectrum.

\section{Summary and Conclusion}\label{sec6}
Recently, NANOGrav and several other pulsar timing array experiments (EPTA, IPTA, PPTA, InPTA, and CPTA) have provided evidence for a common red-spectrum stochastic signal across pulsars in the low-frequency regime. In addition, compelling evidence for the Hellings-Downs inter-pulsar correlation has been observed, which provides the first detection of the stochastic gravitational wave background in the nHz frequency range. There are several possible explanations for the origin of such an SGWB, for instance, a background from merging supermassive black hole binaries, cosmological origins, and so on. Testing whether the cosmological explanation is consistent with the observed data is essential, as it helps discriminate among these candidate origins. In this work, we test whether the SGWB has an inflationary origin, and use the resulting constraints on the primordial GW spectrum to bound the parameters of the inflationary reheating phase.

The primary result of our analysis, shown in Figure~\ref{triangular}, is that the NANOGrav data prefer an extremely blue-tilted spectrum with a large equation-of-state parameter for reheating. In particular, by performing an MCMC analysis we have obtained $n_t=2.94^{+0.24}_{-1.9}$, $\omega_{\text{re}}=0.26^{+0.16}_{-0.43}$ in addition to placing an lower bound on tensor-to-scalar ratio, $r>2.51\times 10^{-15}$ for fixed reheating temperature, $T_{\text{re}}=50$ MeV. Our fit shows that both matter-like ($\omega_{\text{re}}=0$) and radiation-like ($\omega_{\text{re}}=1/3$) reheating scenarios are viable at $68\%$ CL.

Generally, a majority of the studies on primordial GWs sourced by tensor perturbations assume that the inflationary vacuum is a  Bunch-Davies vacuum. However, there is no solid evidence for such a choice. 
In this work, we explore a two-parameter Bogoliubov family of non-Bunch-Davies primordial vacuum and examine the NANOGrav observations, suggesting that the SGWB has an inflationary origin. We find that the NANOGrav 15-year data constrain the parameters $\alpha$ and $\beta$ that characterize the particular subclass of non-Bunch-Davies vacuum. Notably, observations from CMB, particularly the small amplitude of the scalar power spectrum and the tensor-to-scalar ratio, indicate that $\beta_t$ has to be zero. Therefore, within the adopted framework, the NANOGrav observations favour a specific type of vacuum, namely the alpha-vacuum parametrized by $\alpha_t$. Furthermore, results of an MCMC analysis depicted in Figure \ref{triangular2} show that the NANOGrav data significantly narrow the range of the parameter $\alpha_t$ by reducing its upper limit to $\alpha_t<16.5$. For the first time, we provide a phenomenological constraint on the parameters that define the alpha-vacuum. Additionally, we note that, within the framework of a non-Bunch-Davies vacuum, a matter-like reheating scenario requires larger $\alpha_t$, which are incompatible with the NANOGrav constraints and violate the backreaction bound. Therefore, we conclude that the matter-like reheating scenario is disfavoured by NANOGrav observations for a non-Bunch-Davies vacuum.

We have previously noted that observation requires $n_t>0$; as a result, the GW spectrum exceeds the upper limit set by the Planck collaboration when extrapolated to higher frequencies. This issue is common across all scenarios that attempt to explain the observed SGWB as having an inflationary origin. In our setup, we have found that if the parameter $\alpha_t$, which describes the non-Bunch-Davies vacuum, exhibits a specific frequency dependence above a threshold, the blue-tilted issue can be resolved. Therefore, with a suitable choice of the rate of change of $\alpha_{t}$ beyond this threshold frequency, the GW spectrum begins to decline. Moreover, we find that a frequency-dependent parametrization of $\alpha_t$ beyond a specific frequency provides the simplest approach to resolve the long-standing blue-tilted issue. Such frequency-dependent parametrization can also affect the scalar power spectra by suppressing them on small scales. Finally, we have shown that the resulting GW spectrum will be further tested in future GW experiments, including LISA, DECIGO, Cosmic Explorer, and the Einstein Telescope.

\acknowledgments
The authors would like to thank Nilay Kundu and Apratim Kaviraj for discussions and insightful comments. DC acknowledges funding from the ANRF, Government of India, under grant ANRF/CRG/2021/007579. RN would like to thank the MHRD, Government of India for the research fellowship. SS is supported by NPDF grant PDF/2023/002076 from the Science and Engineering Research Board (SERB), Government of India. DC and SS also acknowledge support from an initiation grant IITK/PHY/2019413 at IIT Kanpur and funding from the Indian Space Research Organisation (ISRO) under grant STC/PHY/2024427Q.

\appendix
\section{{Backreaction constraint on $\alpha_t$}}\label{backreaction}
In this appendix, we briefly discuss possible backreaction constraints on the non-Bunch-Davis vacuum following Ref.~\cite{Cielo:2024poz,Tanaka:2000jw,PhysRevD.66.123510,Brandenberger:2012aj}.
A departure from the standard Bunch-Davies vacuum leads to an excited initial state with a non-vanishing particle occupation number. The associated energy density may backreact on the inflationary background, thereby spoiling the inflationary dynamics. Thus, its contribution to the total energy density must remain sufficiently suppressed relative to the background energy density that drives inflation. The energy density associated with the produced particles is given by 
\begin{equation}
    \varepsilon \sim \frac{1}{a^4}\int_0^{a\Lambda}\; \frac{d^{3}k}{(2\pi)^{3}}\; k\; \left|B\right|^2 \; ,  
\end{equation}
where $\Lambda$ denotes the physical UV cut-off, $|B|^{2}$ is the occupation number of the excited states above the Bunch-Davis vacuum, and $B$ is the Bogoliubov coefficient introduced in Eq.~(\ref{NBD}). In our analysis, we adopt the Mottola-Allen vacuum for which $B=e^{i\beta}\sinh(\alpha)$. Consequently, the energy density $\varepsilon$ during the inflationary epoch scales as $\sim \sinh^2(\alpha)\Lambda^4$. Denoting the characteristic scale of the inflationary potential by $V_0$, the requirement that the energy density of these excited states do not spoil the inflationary dynamics is therefore
\begin{equation}
    \sinh^2(\alpha)\Lambda^4 \ll V_0.
    \label{back_cons}
\end{equation}
During inflation, the universe's background energy density is dominated by the inflaton potential energy density. Consequently, the Friedmann equation yields $V_0 \simeq 3\, M_P^2\, H_{\rm inf}^2.$ Taking $M_P=2.435\times10^{18}$ GeV and adopting the benchmark value $H_{\rm inf}=3\times10^{13}$ GeV, we obtain
\begin{equation}
V_0 \sim 10^{64}~\mathrm{GeV}^4.
\end{equation}
For a UV cut-off, $\Lambda \sim 10^{14}$ GeV, the backreaction constraint in Eq.~(\ref{back_cons}) reduces to
\begin{equation}
\sinh^2(\alpha)\ll10^{8}.
\label{back_cons_sim}
\end{equation}
In our analysis, we choose $\alpha_t=8.0$, for which $\sinh^2(\alpha)\simeq 2.2\times 10^{6}$. Furthermore, the chosen value of $\alpha_t=8.0$ is consistent with the upper bound shown in Figure \ref{triangular2}, which is obtained for $H_{*}/M_P = 2\times 10^{-5}$. Choosing a smaller value of $H_{*}/M_P$ results in a further reduction in the upper bound on $\alpha_t$, which indicates that the backreaction constraint is even easier to satisfy.

\bibliographystyle{JHEP}
\bibliography{ref}
\end{document}